\documentclass[12pt,preprint]{aastex}

\shorttitle{StarScan Plate Measuring Machine}
\shortauthors{Zacharias et al.}

\begin{document}

\title{The StarScan plate measuring machine: overview and calibrations}


\author{N.~Zacharias  
      , L.~Winter\altaffilmark{1} 
      , E.~R.~Holdenried\altaffilmark{2}  
      , J.-P.~De Cuyper\altaffilmark{3}
      , T.~J.~Rafferty\altaffilmark{2}
      , G.~L.~Wycoff }
\affil{U.S.~Naval Observatory, 3450 Mass.Ave.NW, Washington, DC 20392,
      \email{nz@usno.navy.mil} }

\altaffiltext{1}{contractor, Hamburg, Germany}
\altaffiltext{2}{USNO, retired}
\altaffiltext{3}{Royal Observatory of Belgium, Uccle, Belgium}

\begin{abstract}
The StarScan machine at the U.S.~Naval Observatory (USNO)
completed measuring photographic astrograph plates to allow
determination of proper motions for the USNO CCD Astrograph
Catalog (UCAC) program.
All applicable 1940 AGK2 plates, about 2200 Hamburg Zone 
Astrograph plates, 900 Black Birch (USNO Twin Astrograph) 
plates, and 300 Lick Astrograph plates have been measured.
StarScan comprises of a CCD camera, telecentric lens,
air-bearing granite table, stepper motor screws, and Heidenhain 
scales to operate in a step-stare mode. 
The repeatability of StarScan measures is about 0.2 $\mu$m.
The CCD mapping as well as the global table coordinate
system has been calibrated using a special dot calibration plate
and the overall accuracy of StarScan $x,y$ data is derived to be 
0.5 $\mu$m.
Application to real photographic plate data shows that position 
information of at least 0.65 $\mu$m accuracy can be extracted from
course grain 103a-type emulsion astrometric plates.
Transformations between ``direct" and ``reverse" measures
of fine grain emulsion plate measures are obtained on the 0.3 $\mu$m
level per well exposed stellar image and coordinate, which is
at the limit of the StarScan machine.
\end{abstract}



\section{Introduction}

This paper introduces the current StarScan hardware, and gives 
details about its operation, calibration, and its scientific program.
This effort is part of the IAU 
Task Force on Preservation and Digitization of Photographic Plates 
(PDPP) \footnote{www.lizardhollow.net/PDPP.htm} to measure astronomical 
plates worldwide.
Reductions of these astrograph data is in progress and astrometric
results will be presented in an upcoming paper.

The StarScan machine in its original set-up was acquired by
the U.S.~Naval Observatory (USNO), Washington in the 1970s.
A projection system \citep{tac1} was used to determine
centroids of stellar images in a semi-automatic mode
using a list of target stars to be measured.
The Twin Astrograph Catalog (TAC) based on over 4000 plates
resulted from these measures \citep{tac1, tac2}.
In the early 1990's StarScann operations stopped and only due to its
extremely large and heavy granite table the machine was not
thrown out, instead was sealed off in a small room
in its original building with a new access door to the outside.

Starting in 1998 \citep{baas} StarScan was completely 
refurbished switching to the current operation mode with an 
imaging CCD camera and step-stare scanning of entire plates.  
The hardware change was performed by L.~Winter (contractor), 
T.~Rafferty (USNO, now retired) and the USNO instrument shop.
The new operations software was developed by E.~Holdenried
(USNO, now retired), and the raw data reduction pipeline was
adopted from a similar program at the Hamburg Observatory
(L.~Winter).

In section 2 of this paper the current science program is described, 
aiming at early epoch star positions to derive proper motions.
Section 3 presents details on the hardware including the 2006 
upgrade and section 4 describes the calibration process from set-up 
of dedicated measures to interpretation of results.
Section 5 reveals external errors derived from measures
of real astrometric plates, and section 6 summarizes the
astrometric performance of StarScan.

\section{Science Program}

The current StarScan operation is part of the USNO CCD Astrograph
Catalog (UCAC) program \citep{ucac1,ucac2}.
All-sky, astrometric observations were completed in 2004
and reductions for the final release are in progress.
StarScan will provide early epoch positions on the Hipparcos
system \citep{hip} from these new, yet unpublished measures.
The UCAC program aims at a global densification of the
optical reference frame to magnitude R=16 including
proper motions for galactic dynamics studies.
A variety of early epoch data are used for UCAC with the
StarScan measures allowing an improvement over Tycho-2 \citep{tyc2}
and extending to fainter magnitudes at a high accuracy level
for a large fraction of the sky.

Table 1 lists the plates in the just completed StarScan measure
program.  Most plates were taken with dedicated,
astrometric astrographs of about 2 meter focal length.
AGK2 is the {\em Astronomische Gesellschaft Katalog} data
taken at Hamburg and Bonn observatories between 1929 and 1933,
covering completely the $-2.5^{\circ}$ to $+90^{\circ}$ 
declination range with a blue limiting magnitude of about 12.
In a 2-fold overlap pattern of plates each area of the sky in
that declination range is observed on 2 different plates.
Only a small fraction of the stars visible on the plates could 
be measured and reduced in a several decades long, manual program 
to produce the published AGK2 catalog \citep{agk2}.
The data were then combined with the 1960's re-observations
to form the AGK3 which includes proper motions based on 
about 30 years epoch difference between AGK2 and AGK3 plates.
Combining the new measures of the 1930 epoch AGK2 data with
the UCAC CCD survey at epoch $\approx$ 2001 will result in
proper motions accurate to about 1 mas/yr, thus significantly
improving upon the Tycho-2 result.

Between 2002 and 2003 all 1,940 applicable, original AGK2
plates were completely digitized on StarScan.
This re-measure effort was carefully planned and prepared
by the late Christian de Vegt (1978), Hamburg Observatory.
About 40\% of these new measures entered the UCAC2 release
from preliminary reductions available at the cut-off date.
The complete AGK2 data will enter the UCAC3 (in preparation).

The BY, ZA, and Lick columns in Table 1 give details of the
Black Birch (USNO Twin Astrograph, New Zealand) Yellow Lens, 
the Hamburg Zone Astrograph (ZA), and the Lick Astrograph plates, 
respectively.
Most of these BY and ZA plates were taken around
radio sources (quasars and radio stars) used in the 
International Celestial Reference Frame (ICRF), 
and together they cover about 35\% of the sky
(both hemispheres).  The limiting magnitude is about V=14.
When combined with the more recent CCD observations,
proper motions on the 3 to 5 mas/yr level can be derived for 
millions of stars beyond the Tycho-2 limiting magnitude.
The Lick astrograph has a more favorable image scale
than BY and ZA, and also uses mostly fine grain emulsions.
However, only a limited number of fields were observed
during this decade long test program at an epoch around 1990.

\section{StarScan Details}

\subsection{Hardware}

Figure 1 shows the StarScan machine as of July 2006.
The granite base measures 2.4 by 1.5 by 0.3 meter,
and plates up to 260 mm square can be measured.
An $x,y$-table moves on the base via pressure air bearings
while a CCD camera is mounted on a granite cross-beam 
which is fixed to the base.

A photographic plate is put inside a plate holder on a rotation disk
which is mounted on the $x,y$-table.
With emulsion up, the plate is clamped upwards against a fixed mechanical 
reference plane.  Thus the location of the plate emulsion is always
at the same distance from the lens and no change in focus is required 
during operations even if different types of plates are being measured.
For very thin plates a supporting clear class plate can be inserted 
underneath the plate to be measured.
The $x,y$-table is moved by precision screws and stepper
motors while the coordinates are read from Zerodur Heidenhain
scales to better than 0.1 $\mu$m precision.

The plate is illuminated from below with a diffuse light projection
system and a 2-sided telecentric objective (Schneider Xenopla, imaging 
scale 1:1) maps a section of the plate being measured onto a CCD.
From the year 2000 until spring 2006 a Pulnix CCD camera
with 1280 by 1030 pixels of 6.7 $\mu$m size was used.
Thereafter a 2k by 2k QImaging camera with 7.4 $\mu$m
square pixels (Kodak chip) is used.
Figure 2 shows a close-up view of the area with the plate
and the new CCD camera attached to the lens.

The new camera was selected to be compatible with the
existing system in order to minimize changes.
However, the upgrade to the new camera did not go 
smoothly due to various interface issues and in the end 
a change of the operating system and a complete re-write
of large portions of the code was required as well
as a new mechanical adapter.

\subsection{Measure Process}

Most functions, including disk rotation and its clamping
mechanism are under computer control.
At the beginning of each plate measure the illumination
is manually set to just below saturation on the plate
background by adjusting the power supply unit to the lamp.
Then the plate is measured automatically in step-stare mode.
The $x,y$-table moves to the next grid coordinates,
the actual table coordinates are obtained by reading out
the calibrated Heidenhain scales,
a digital image is taken, and another read out of the
actual table coordinates is performed.
If the table coordinates from the 2 readings
differ by more than 0.1 $\mu$m the procedure is repeated.
Meanwhile the image is processed: bias, flat field,
and dark current corrections, object detection and image profile fits.
The entire cycle took about 2.5 sec with the old camera
and takes about 4.0 sec with the new camera, which covers
about 3 times more area, resulting in an overall
throughput gain of a factor of 2.

A photographic plate is thus digitized to 10 bit with
a resolution of now $7.4 \mu$m per pixel.
Adjacent pictures have overlap.  The usable field-of-view
of the old camera was 8.71 by 6.86 mm, while grid steps of
about 7.5 and 5.9 mm for $x$ and $y$, respectively, were used.
The new camera has a usable field of about 13 by 13 mm,
with slightly degraded image quality (field curvature)
visible in the corners.  A grid step size of about 11 mm
is adopted between adjacent images.

After the plate measure is complete, the disk is unclamped,
rotated by $180^{\circ}$, clamped, and the ``reverse" measure
started.  With the new set-up a measure of a single plate in 
2 orientations takes about 45 minutes.
All pixel data (about 3.6 GB compressed per plate) are
transferred overnight and saved to DVD the following day.
The stellar image profile fit results are collected on
hard disk for further processing.

\subsection{Notes on Reductions}

The bias, flat field and dark current corrected pixel data of
the CCD camera are used in 2-dimensional fits to obtain centroid
coordinates of stellar images.  A double exponential model
function is used for this purpose \citep{lars_phd},


\large

\[ I(x,y) = B \ + \ A \ 
   ( 1 - e^{-ln2 \ e^{-\frac{8}{ln2} \ S \ (r - r_{0})}} ) \]

\[  r = \sqrt{(x - x_{0})^{2} + (y - y_{0})^{2}}  \]

\normalsize
with the 6 independent parameters $B$ = background intensity level, 
$A$ = amplitude,
$x_{0}, y_{0}$ = image centroid coordinates, $r_{0}$ = radius
of profile, and $S$ = shape parameter.
This function models 
the shape of the observed profiles sufficiently well for
all levels of brightness of the stars, while e.g.~a simple
Gaussian function does not match the saturated bright star
profiles with ``flat tops" very well.
Due to the fine sampling (at least 4 pixels across the
diameter of the smallest stellar images) a sufficiently large
number of pixels contain signal to determine these 6 profile
parameters per stellar image.
The internal fit precision for all well exposed stars is
typically in the order of 0.1 $\mu$m or even below for
relatively bright, circular symmetric stellar profiles.
For quality control, the internal image profile fit errors 
are plotted versus the instrumental magnitude for all stellar 
images of a single plate measure and orientation.

In order to arrive at highly accurate, global $x,y$ coordinates
of stellar images from these plate measures several assumptions
need to be verified and calibration parameters need to be obtained.
Only the center of the CCD chip (or some other, fixed, adopted
reference point) is initially associated with the $x,y$-table
coordinate readings from the scales.
The positions of arbitrary stellar images measured on the
CCD chip field-of-view in pixel coordinates need to be referred 
to the global, $x,y$-table coordinates of the machine, i.e.~the
mapping model and parameters need to be obtained for this
transformation.  An important parameter here is the third order
optical distortion of the telecentric lens.
Using the dot calibration plate (see below) StarScan has
been set up to give a mapping scale of 1.0 to within less
than a percent and the CCD pixel coordinates are aligned
to the global $x,y$-table coordinates to within $10^{-4}$ radian.
Also geometric distortions of the observed, global,
$x,y$-table coordinates with respect to ideal rectangular
coordinates need to be investigated.
For both steps, calibrations of the StarScan machine
are being made (see below).

Finally, the ``direct" and the ``reverse" measures of the
same plate are compared to check for possible systematic 
errors e.g.~as a function of magnitude or $x,y$ coordinates
introduced by the measuring process and
to verify that the applied mapping procedures were adequate.
The direct-minus-reverse residuals also identified several
unexplained glitches of some plate measures, which could 
have been caused by thermal drifts or electronic problems.
The set of the ``reverse" measures $X, Y$ (after rotation 
by $180^{\circ}$) is fit to the ``direct" measures ($x, y$)
with the following transformation model,

\[ X \ = \ a \ x \ + \ b \ y \ + \ c \ + \ e \ x \ + \ f \ y 
            \ + p \ x^{2} \ + \ q \ x y \]

\[ Y \ = \ -b \ x \ + \ a \ y \ + \ d \ + \ e \ y \ - \ f \ x 
            \ + q \ y^{2} \ + \ p \ x y \]

where $a,b,c,d$ are the parameters for the orthogonal model,
$e,f$ the non-orthogonal, linear parameters, and $p,q$ account
for a possible tilt between the 2 sets of measures.
On the precision level investigated here the rotation axis
of the disk is not perfectly perpendicular to the plane of
the plate and the $p,q$ terms are required.
These tilt terms are reproducible to some extent,
however, sometimes a grain of dust would be trapped between the
plate and reference plane of the machine, resulting in
significantly different tilt terms.
It is important to realize that even after applying the above 
transformation (and averaging of the $x, y$ coordinates from the 
``direct" and the `rotated ``reverse" measures) the data set is 
affected by an unknown tilt, the tilt of the ``direct" measure 
photographic plate surface with respect to the plane of the 
$x,y$-table motion. 
Thus for the astrometric reductions to follow (from $x,y$ to 
$\alpha, \delta$) the minimal plate model has to include tilt 
terms as well, even if the tangential point would be known 
precisely and in the absence of a tilt between the optical axis 
of the telescope and the photographic plate.

\section{Calibration}

\subsection{Dot calibration plate}

The Royal Observatory of Belgium bought a dot calibration plate (DCP)
which was made to specifications for the purpose of calibrating
photographic plate measuring machines.
This 251 mm by 251 mm by 4.6 mm glass plate is highly transparent
except for circular, opaque, metal dots of $\approx$ 50 to 
300 $\mu$m diameter which are put on the plate in a regular grid
covering the inner 240 mm by 240 mm.
Medium size dots (100 $\mu$m) occupy a regular grid of 1.0 mm spacing
with small dots (50 $\mu$m) put in between every 0.5 mm along both 
coordinates. 
In hierarchical order the dots are slightly larger for 5 mm,
10 mm, 50 mm, and 100 mm major grid lines and located at the line
intersections of such.
The orientation of the grating is uniquely marked by a few
out-of-symmetry dots and a missing small dot every cm.

The dot calibration plate was manufactured by silicon wafer
technology machines and the location of the dots are believed
to be accurate to within about 0.1 to 0.2 $\mu$m  for both
the small-scale regularity and the global, absolute coordinates
over the entire area of the plate.

The dots resemble very high-contrast, stellar images and
are processed with the standard reduction pipeline at StarScan.
The internal position fit errors of these dots are typically 
0.1 $\mu$m for the medium dots on the 1 mm grid, slightly larger for 
the small 0.5 mm grid dots and even smaller for the major grid dots.

\subsection{CCD camera mapping}

Initial StarScan mapping parameters for the CCD pixel 
coordinates to global $x,y$-table coordinates transformation
are determined from repeated measures of the dot calibration plate, 
centered on main grid lines and dithered by small offsets in $x$ 
and $y$.  This in particular allows the determination of the 3rd order
optical distortion term $D$ of the lens, which is then used in the 
preliminary pipeline, together with orientation and scale information,
to identify the same star as seen in 2 or more 
adjacent, overlapping CCD images.
Figure 3 shows the residuals of the 2k camera optical distortion 
pattern after removal of that 3rd order term.  The working area
used for plate measuring is about 11 by 11 mm.  The scale of the
residual vectors is 1000, thus the largest residual is about 
1.0 $\mu$m, with most of the residuals in the 0.1 to 0.2 $\mu$m
range.  
The point spread function (PSF) of images in the CCD working area
is uniform, and there is no variation of the PSF over the area of
the plate because there is no variation of best focus over the
plate area due to the hardware setup described above.

It is assumed that the mapping parameters stay constant for each
individual plate measured in a single orientation, thus all
overlap images of such a dataset are used simultaneously
to determine the following 7 mapping parameters $a,b, e,f, p,q, D$ 
from

\[ \Delta X \ = \ a \ x \ + \ b \ y \ + \ e \ x \ + \ f \ y 
            \ + p \ x^{2} \ + \ q \ x y \ + \ D \ x (x^{2} + y^{2})\]

\[ \Delta Y \ = \ -b \ x \ + \ a \ y \ + \ e \ y \ - \ f \ x 
            \ + q \ y^{2} \ + \ p \ x y \ + \ D \ y (x^{2} + y^{2})\]

where $\Delta X, \Delta Y$ are the coordinates of a star
in the global, $x,y$-table system (in millimeter) with respect 
to the current position as read by the scales, $x, y$ are the
stellar image coordinates on the CCD (in pixel) with
respect to the center of the CCD.
The difference between 2 such equations per coordinate is 
observed for each pair of overlapping images on 2 CCD frames
of the same star.

There are no constant terms because the zero point of this
transformation is arbitrary, thus there are only 4 linear terms
$a,b,e,f$.
The $p,q$ terms represent the tilt between the plane of 
the CCD chip inside the camera and the photographic plate.
The higher order terms for distortion and tilt are almost
constant for a large number of plates of the same type
and a mean value can be determined and used as fixed parameter
in the CCD mapping procedure, while solving then for the
linear parameters for each individual plate and orientation
measure set with significantly lower errors.
Results will be presented in an upcoming paper of this series.

\subsection{DCP measures}

The DCP was measured in 4 orientations (000, 090, 180, 270
degrees) consecutively with a $x,y$-table step size of
5 mm, major dot grid lines well centered, and grid lines
precisely aligned to the table motion.
The entire set of measures was repeated about 2 weeks later,
with many regular survey plates measured in between.
For the present purpose only the {\em central dot} of each
CCD image has been evaluated which fell on the same
spot on the CCD chip within about $\pm$ 10 pixels. 
Thus any errors introduced by the CCD mapping model are 
negligible and we measure here the accuracy of the
$x,y$-table coordinates of StarScan.
In the following we refer to these sets of measures as 
{\em m1c} and {\em m2c}, each consisting of 4 single
orientation measures of 2025 dots each (on a 220 mm square
area with 5 mm step size).

\subsection{Repeatability}

Without any assumptions on the errors of the DCP
a comparison between 2 measures of the same dots in the
same measure orientation reveals the absolute repeatability 
of StarScan measures.  Using a linear transformation model 
between the 2 sets of measures of the 2025 dots gives 
the results in line 1 through 4 in Table 2 for the root-mean-square
(RMS) observed differences per coordinate.
The mean of the 4 orientations is presented on line 5 and 
example 2-dim vector plots of the measured position differences
are shown in Figure 4.
All vectors are small ($\approx$ 0.3 $\mu$m) but not randomly
oriented.  For parts of some lines along $x$ they are highly 
correlated, mainly affecting the $\Delta x$ coordinate.
Examples are seen in Figure 4 on the left hand side (for
the $90^{\circ}$ orientation case) for $y$ = $-$110, $-$100,
$-$30, +40, and +60 mm, and for the $180^{\circ}$ case
(right hand side of Figure 4) for $y$ = $-$70, +30, and +50 mm.
These glitches are currently not well understood,
and seem to appear randomly for a typically length of 50 to 
150 mm along the $x$ axis.
Some of these errors will cancel out when combining the
``direct" and ``reverse" measure of a plate.
These ``glitches" were not further investigated and currently
represent the physical limit of the StarScan measure accuracy.

To arrive at the repeatability error ($\sigma_{rep}$)
of a single StarScan measure over the entire plate area and
per coordinate we divide the RMS differences (Table 2, result
line 5, showing the scatter between 2 measures) 
by $\sqrt(2)$ and thus obtain 0.23 and 0.15 $\mu$m
for the $x$ and $y$ coordinate, respectively.
This error is not dominated by the fit precision of the
dots as explained above, instead shows the limit of accuracy 
of the $x,y$-table coordinate measures.

\subsection{Corrections to $x,y$-table coordinates}

Next we investigate systematic errors of the $x,y$-table
coordinates, i.e.~the global, geometric distortions.  
For each data set (2 measures, each with 4 orientations) a 
least-squares fit is performed of the measured 
coordinates to the nominal, ideal grid coordinates, using
the above mentioned 8-parameter model which includes
linear and tilt terms.
Results are presented in Table 2, lines 6 to 13 and for some
cases plots are shown in Figure 5.
As can be seen the difference vectors are highly correlated
among the different plots, indicating that the dominating
error source is systematics in the StarScan $x,y$-table geometry,
due to imperfections of the axes as the plate is moved along
in the measuring process.
If the DCP would have large errors as compared to StarScan,
the rotated measures ($90^{\circ}$ and $180^{\circ}$) would
show the same pattern as the ``direct" measure (orientation 
$0^{\circ}$) but rotated by that angle.
However, we see here the systematic error pattern correlated
directly to the StarScan $x,y$ coordinates, regardless of the
orientation of the calibration plate.

A mean 2-dimensional vector difference map was constructed
by averaging all 8 measures (2 sets times 4 orientations each)
and then applying a near neighbor smoothing.
Results are given on line 14 in Table 2 and the corresponding
vector plot is shown in Figure 6.
Subtracting this mean pattern from the individual measure
sets (Table 2, lines 6 to 13) results in lines 15 to 22
with RMS values almost as low as the StarScan repeatability,
showing the successful removal of most of the $x,y$-table 
systematic errors.

\section{Application to Real Data}

\subsection{Different plates of the same field}

In order to determine the physical limit of astrograph
plates for astrometry a set of 4 different plates taken of
the same field (2234+282) in the sky is investigated.
The 4 plates were taken in 1988 with the Hamburg Zone Astrograph
(ZA) \citep{za_cdv} in the same night within 40 minutes off the
meridian through an OG515 filter on 103aG emulsion.
Plate numbers 1929 and 1930 were taken with the telescope
on the West side of the pier, numbers 1931 and 1932 on East.
Here we assume that the StarScan measure errors are small
and we will try to determine total external errors from the
combination of plate geometry (emulsion shifts) over the entire 
field of view and the background noise from the emulsion grains,
using real stellar images.

A transformation of the ``direct" measure of ZA1930 onto
the ``direct" measure of ZA1929 was performed using the
above mentioned 8-parameter model and using about 19,000 stellar
images in an unweighted, least-squares fit.
A 5-parameter mapping model including linear and tilt
terms was used for the CCD camera mapping, after applying
a mean optical distortion term to the raw data.
No $x,y$-table corrections of StarScan were applied because
they would cancel anyway in this ``direct" to ``direct"
transformation (assuming stability of the machine over a
few hours in which the set of plates were measured). 
Table 2, lines 23 to 26 show the mean RMS differences for
``well exposed" stars (about 10th magnitude) between the
2 plates.
Thus the external error of a single, well exposed stellar
image on a single plate is about 0.65 $\mu$m per coordinate.

This number reflects the sum of the StarScan measure errors
and the errors inherent in the photographic material,
and is dominated by the grain noise of this type of emulsion,
consistent with previous investigations \citep{vanA}.
However, these numbers do not directly translate into $\alpha, \delta$
accuracies.  In order to get highly accurate astrometric positions on
the sky from these calibrated, measured $x,y$ data the systematic
errors introduced by the astrograph (optics, guiding etc.) and the 
atmosphere need to be detected, quantified and removed.

\subsection{Long term repeatability and discussion}

A fine grain (Kodak Technical Pan emulsion) plate (No.~724) of a
dense field (NGC 6791) obtained at the Lick 50cm astrograph has
been measured multiple times during the StarScan operations
at relatively regular intervals to monitor its performance.
Contrary to the DCP, the location of the ``dots" on this
plate are not known {\em apriori}; however, they are ``fixed"
and StarScan should measure the same positions of these
stellar images (within its errors) every time it has been
measured. Thus these data sets provide information about the
long-term stability and repeatability of the StarScan measures
and shed some light on the validity of using ``the" $x,y$-table
correction pattern as derived in the previous section.

Results are given in Table 2 for measures 1,6,9, and ``a" 
(lines 27 to 30), which correspond to May 2004, Sept 2004,
Sep 2005, and Apr 2006, respectively.
For each data set the plate was measured in ''direct" (000)
and ``reverse" (180) orientation and the 8-parameter model
was used in a least-squares adjustment in the transformation
between the orientations $x,y$ data after applying the same 
2-dim table corrections as presented above to all measures.

Figure 7 shows the residuals of such a transformation for
the measure \# 6 (September 2004) of that Lick plate. 
As an example the $x$ residuals are shown as a function of
$x$, $y$, and magnitude.
There is a magnitude equation of about 0.3 $\mu$m amplitude
for $x$,
while there is none for the $y$ coordinate (not shown here).
Residuals as a function of the coordinates still show
systematic errors up to about 0.5 $\mu$m, even after applying
the 2-dim $x,y$ table corrections.
The high frequency wiggles are likely caused by the residual
CCD camera mapping errors, while the low frequency systematic
errors are likely deviations of the $x,y$ table corrections
as applied with respect to what actually is inherent in
the data at the time of the measurement.

In order to better illustrate the long-term variability,
Figure 8 shows only the $x$ residual versus $y$ plot but
for 3 other measures of the same plate (May 2004, September 
2005, and April 2006), which should be compared to the
middle diagram of Figure 7.
Clearly there are variations over time, however the overall
correction is good to about 0.4 $\mu$m per coordinate,
with the RMS error being much smaller.

Part of the remaining systematic pattern is clearly correlated 
with the location of stellar images on the CCD area.  This is
caused by an imperfect model in mapping from the CCD pixel
coordinates to the $x,y$ table coordinate system.  This is
not surprising considering that the poorest image quality
of the CCD area (ther corner and border areas) are utilized
to determine those mapping parameters, and an extrapolation 
to the entire CCD area has to be made.  Measuring plates
with significantly larger overlap between adjacent CCD
images would help to even lower those type of errors.
However, as other error sources indicate, we have reached the 
overall StarScan accuracy limit and allowing a significantly
larger overlap would have increased the project time
significantly with little benefit.
The HAM-I measuring machine \cite{ham1} completely eliminated
this problem by centering each star to be measured on the CCD
image.  However, that required an {\em a priori} measure list
and a very long measure time, but an astrometric accuracy
similar to that of StarScan was obtained.

It is not possible to correct the measures for the remaining 
systematic errors as a function of the coordinates based on 
findings like Figures 7 and 8 because the {\em differences} 
between the ``direct" and ``reverse" measure are seen here,
while the combined, mean measured position of each star
would be based on the {\em sum} of the measures.
Comparisons like these for the residuals as a function 
of the coordinates serve as check of the data and applied models,
and are indicative for such systematic errors staying constant
which allows to group plate measures correctly in
further reduction steps.

The situation is much different for the magnitude dependent
systematic errors.
Assuming that the magnitude equation is constant for
both orientation measures, it automatically drops out when
combining the ``direct" and ``reverse" measures, while 
twice the deviation of a single measure with respect to the error
free data is seen in the residual plot as a function of magnitude.
If only a single plate orientation were measured this
magnitude equation would need to be determined and corrected
by some other means, which would be hard to do.

\section{Summary and Conclusions}

For well exposed stellar images StarScan has a repeatability
error of 0.2$\mu$m per coordinate.
Using overlapping areas in the digitization process the mapping
parameters of the CCD camera with respect to the $x,y$-table
coordinate system are calibrated for each plate and orientation
measure to the same precision.
The overall accuracy of StarScan measures is limited by the
systematic errors of the $x,y$-table and the stage motions.
Using a dot calibration plate specifically manufactured for
this purpose, 2-dimensional correction data are obtained to
calibrate the $x,y$-table coordinate system.
The high precision of the StarScan measures allows one to see
variations in this calibration pattern which depend at 
least on time, maybe on other parameters as well.
The uncalibrated machine performs at an accuracy level of
about 1 $\mu$m RMS with a parabola-like distortion for
the $\Delta y$ as a function of $y$ with $\pm$ 1.5 $\mu$m
amplitude being the largest error contribution.
After applying the 2-dim calibration corrections
StarScan performs on the 0.3 $\mu$m per coordinate
accuracy level.

By comparing different photographic plates of the same
field it has been demonstrated that astrometric useful information
at an accuracy level of at least 0.65 $\mu$m can be extracted from
the data of course grain emulsions exposed at dedicated astrographs.
Comparisons of ``direct" and ``reverse" measures of fine grain
emulsions show errors of 0.3 $\mu$m per single star image and
measure.  This is only an upper limit of the plate data quality,
because of StarScan measuring machine errors close to the same level.
A higher accuracy than StarScan currently can provide is
recommended for fine grain astrometric plates and such
a new machine was just delivered to the Royal Observatory
Belgium \citep{d4d}.

It is recommended to measure the dot calibration plate often
to verify and improve the 2-dim $xy$-table corrections.
Measuring a plate in 2 orientations, rotated by $180^{\circ}$,
is mandatory, at least for StarScan and when aiming at high accuracy 
metric results, in order to correct for the non-linear magnitude
equation from the measure process and to monitor quality by
identifying occasional problems in the measure process.



\acknowledgments
The authors thank Sean Urban who oversaw the
operations in the past.
Brian Mason and William Hartkopf are thanked for assisting in
the daily measure effort. 
John Pohlman and Gary Wieder, former and current Head of the USNO
instrument shop, and all instrument shop personnel involved in the
StarScan program are thanked for their effort in maintaining and
upgrading the machine.
Steven Gauss, former head of the Astrometry Department is
thanked for his support of StarScan, particularly in managing to
keep the machine alive. Ralph Gaume, current head of the Astrometry 
Department is thanked for his support, in particular for budgeting 
several contracts which allowed the maintenance and upgrade of StarScan.
Hamburg Observatory (director Prof.~Schmidt) is thanked for
loaning the Zone Astrograph plates to USNO for digitization.

\clearpage

\begin{table}
\begin{center}
\caption{Astrometric plates in the just completed StarScan measure program.
  The AGK2 covers each area of the northern sky twice (2-fold overlap of
  plates).  The BY (south) and ZA (north) programs together cover about
  35\% of the entire sky area.}
\vspace*{5mm}
\begin{tabular}{lcccc}
\tableline
\tableline
project, telescope name      & AGK2     &   BY     &  ZA      &  Lick \\
\tableline
focal length (meter)         &  2.0     &   2.0    &  2.0     &  3.75 \\
plate size (mm)              & 220 sq.  & 200 x 250& 240 sq.  & 240 sq.\\
measurable field size (degree) & 5 x 5  &  5 x 6   &  6 x 6   & 3 x 3 \\
limiting magnitude           & B = 12   &  V = 14  &  V = 14  & V = 15 \\
\tableline
number of plates             &  1940    &   900    & 2200     &  300  \\
range of epochs              & 1929--31 & 1985--90 & 1977--93 & 1980--2000\\
emulsion type                & fine grain & 103aG  & 103aG    &  various \\
declination range (degree)   & $-$2.5 +90 & $-$90 +10 & $-$5 +90& $-$20 +90\\
sky coverage completeness    &  2-fold    & 35\%  &  35\%     &  5\% \\
\tableline
\end{tabular}
\end{center}
\end{table}

\begin{table}
\begin{center}
\caption{Results of transformations between 2 sets of data,
         DCP = dot calibration plate, {\em m1c, m2c} = sets
         1 and 2 of measures of central dots, and 000 to 270 
         indicate the orientation of a measure.
         See text for more details.}
\vspace*{5mm}
\begin{tabular}{cllccc}
\tableline
\tableline
result& data set 1  &  data set 2    &  model   & RMS x  & RMS y  \\
number&             &                &          &($\mu$m)&($\mu$m)\\
\tableline
 1 & DCP m1c 000    & DCP m2c 000    &  linear  &  0.31  & 0.20  \\
 2 & DCP m1c 090    & DCP m2c 090    &  linear  &  0.34  & 0.24  \\
 3 & DCP m1c 180    & DCP m2c 180    &  linear  &  0.30  & 0.15  \\
 4 & DCP m1c 270    & DCP m2c 270    &  linear  &  0.33  & 0.23  \\
\tableline
 5 & \multicolumn{2}{c}{mean of 1--4}&          &  0.32  & 0.21  \\
\tableline
 6 & DCP m1c 000    & ideal grid     &lin.+ tilt&  0.41  & 0.44  \\
 7 & DCP m1c 090    & ideal grid     &lin.+ tilt&  0.47  & 0.53  \\
 8 & DCP m1c 180    & ideal grid     &lin.+ tilt&  0.36  & 0.56  \\
 9 & DCP m1c 270    & ideal grid     &lin.+ tilt&  0.40  & 0.42  \\
10 & DCP m2c 000    & ideal grid     &lin.+ tilt&  0.40  & 0.40  \\
11 & DCP m2c 090    & ideal grid     &lin.+ tilt&  0.44  & 0.57  \\
12 & DCP m2c 180    & ideal grid     &lin.+ tilt&  0.37  & 0.53  \\
13 & DCP m2c 270    & ideal grid     &lin.+ tilt&  0.48  & 0.40  \\
\tableline
14 & \multicolumn{3}{l}{mean, smoothed 2-dim $x,y$-table corrections} &0.27&0.39\\
\tableline
15 & result   6   & mean corrections &  none    &  0.32  & 0.33  \\
16 & result   7   & mean corrections &  none    &  0.35  & 0.23  \\
17 & result   8   & mean corrections &  none    &  0.28  & 0.32  \\
18 & result   9   & mean corrections &  none    &  0.31  & 0.20  \\
19 & result  10   & mean corrections &  none    &  0.28  & 0.32  \\
20 & result  11   & mean corrections &  none    &  0.30  & 0.26  \\
21 & result  12   & mean corrections &  none    &  0.28  & 0.27  \\
22 & result  13   & mean corrections &  none    &  0.37  & 0.27  \\
\tableline
23 & ZA 1929 000  & ZA 1930 000      &lin.+ tilt&  0.89  & 0.85  \\
24 & ZA 1929 180  & ZA 1930 180      &lin.+ tilt&  0.84  & 0.89  \\
25 & ZA 1931 000  & ZA 1932 000      &lin.+ tilt&  0.92  & 0.92  \\
26 & ZA 1931 180  & ZA 1932 180      &lin.+ tilt&  0.92  & 0.92  \\
\tableline
27 & Lick 724 m1 000 & Lick 724 m1 180 &lin.+ tilt& 0.41 & 0.68  \\
28 & Lick 724 m6 000 & Lick 724 m6 180 &lin.+ tilt& 0.40 & 0.35  \\
29 & Lick 724 m9 000 & Lick 724 m9 180 &lin.+ tilt& 0.42 & 0.49  \\
30 & Lick 724 ma 000 & Lick 724 ma 180 &lin.+ tilt& 0.45 & 0.45  \\
\tableline
\end{tabular}
\end{center}
\end{table}


\clearpage

\begin{figure}
\caption{The StarScan plate measuring machine at USNO, Washington DC,
   with Lars Winter on top of things and Gary Wieder in the background.}
\vspace*{5mm}
\plotone{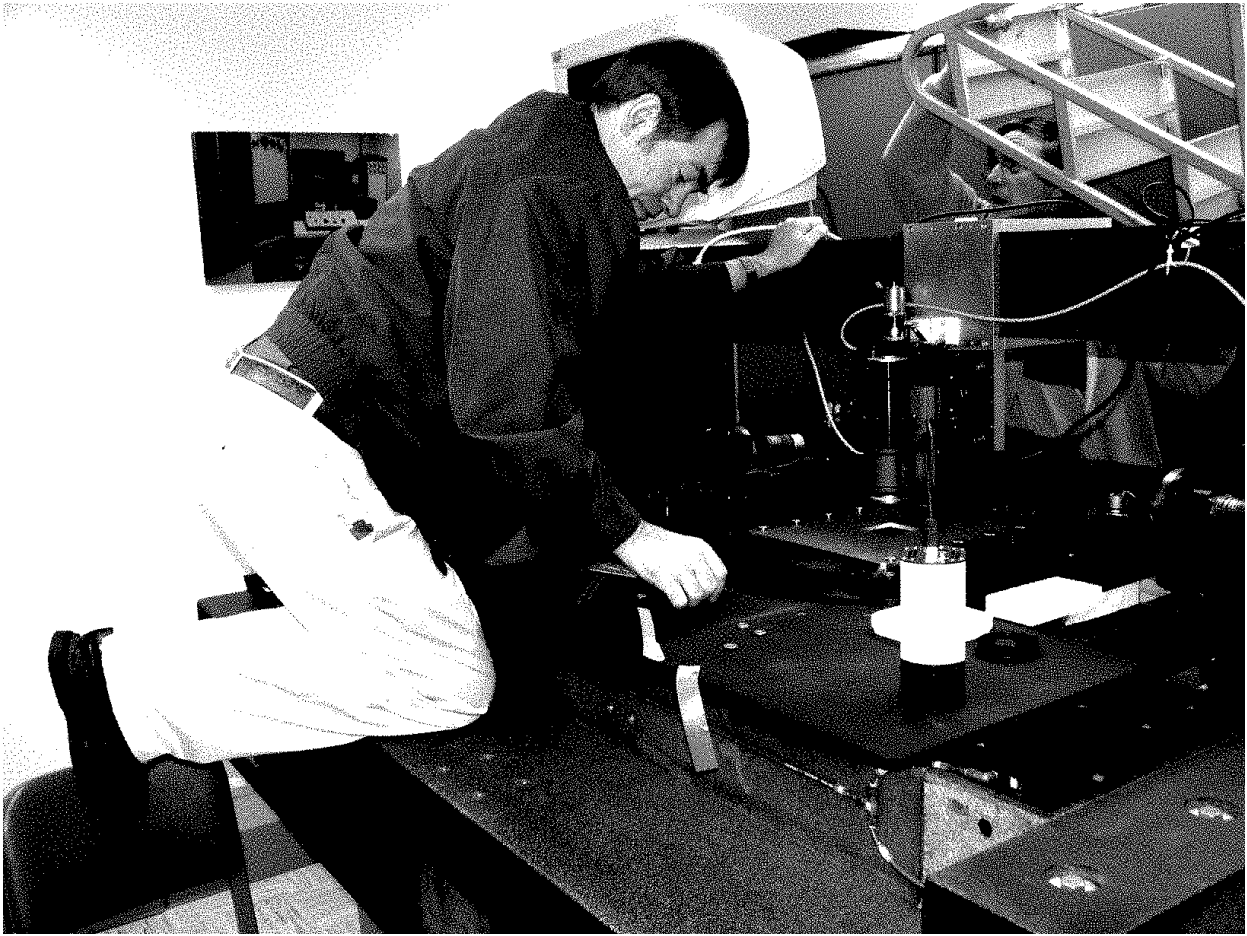}
\end{figure}

\begin{figure}
\caption{Close-up view of the StarScan plate measure area with the 
         2k by 2k QImaging camera and Schneider telecentric lens.}
\vspace*{5mm}
\epsscale{.7}
\plotone{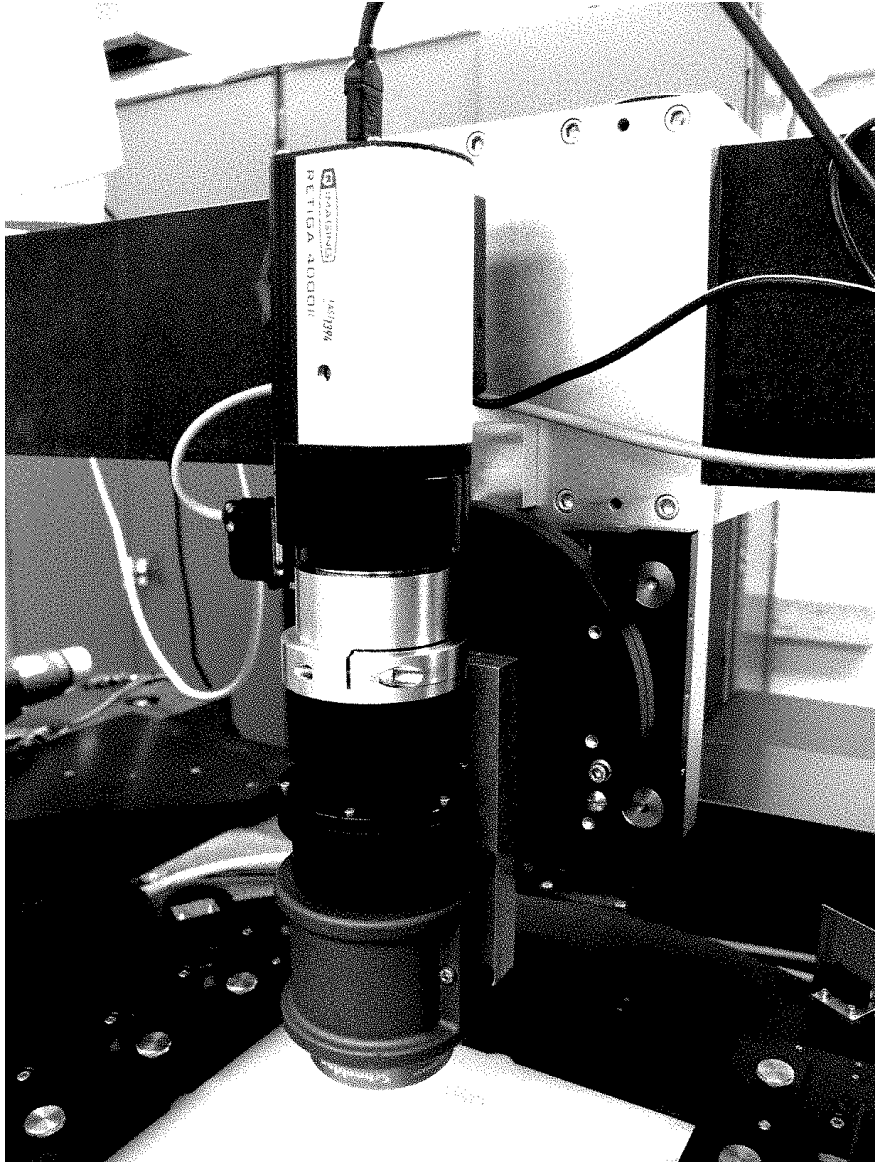}
\epsscale{1}
\end{figure}

\begin{figure}
\caption{Optical distortion pattern of the 2k camera field of
         view after removal of the 3rd order term.
         The residual vectors are scaled by a factor of 1000,
         thus the largest vector is about 1.0 $\mu$m long.
         Results are shown in the about 11 by 11 mm area of the
         detector which is used for plate measuring.} 
\vspace*{10mm}
\includegraphics[angle=270]{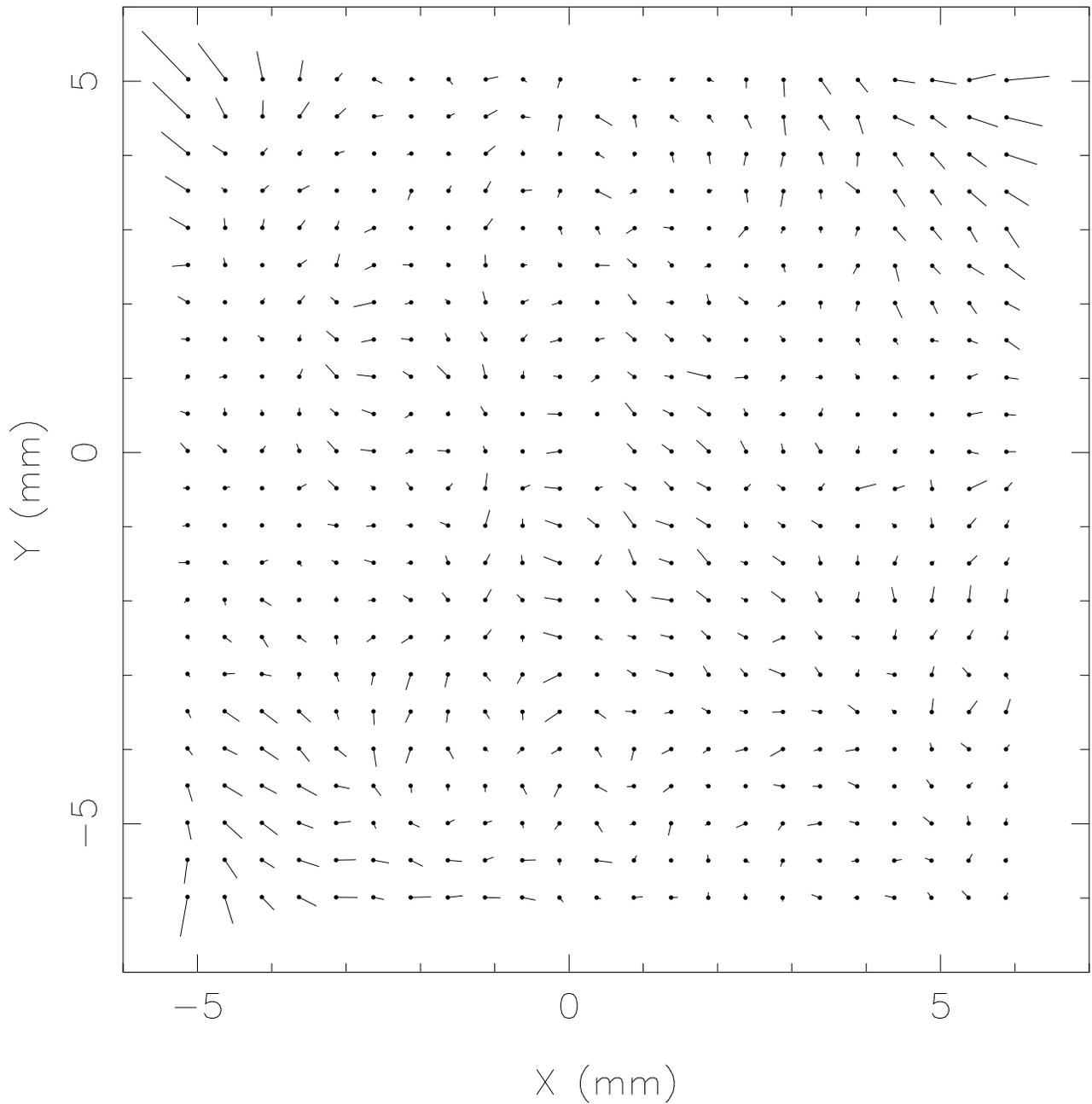}
\end{figure}

\begin{figure}
\caption{Repeatability: vector plot of position differences
         between 2 measures of the same dot calibration plate
         in the same orientation, performed on different days.
         The examples show the worst (left) and best (right) case 
         ($90^{\circ}$ and $180^{\circ}$ orientation, respectively).
         The scale is 5000 thus the largest vectors are about
         1.0 $\mu$m long, and the RMS vector length is 0.42 and
         0.33 $\mu$m, respectively.}
\vspace*{10mm}
\plottwo{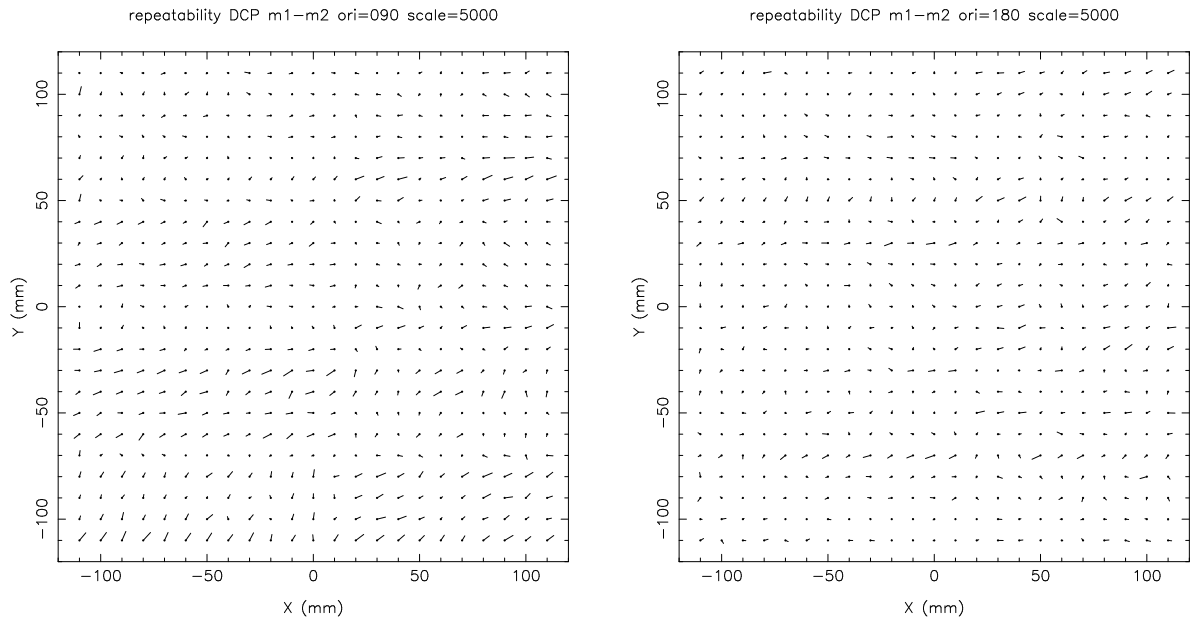}{f4b.eps}
\end{figure}

\begin{figure}
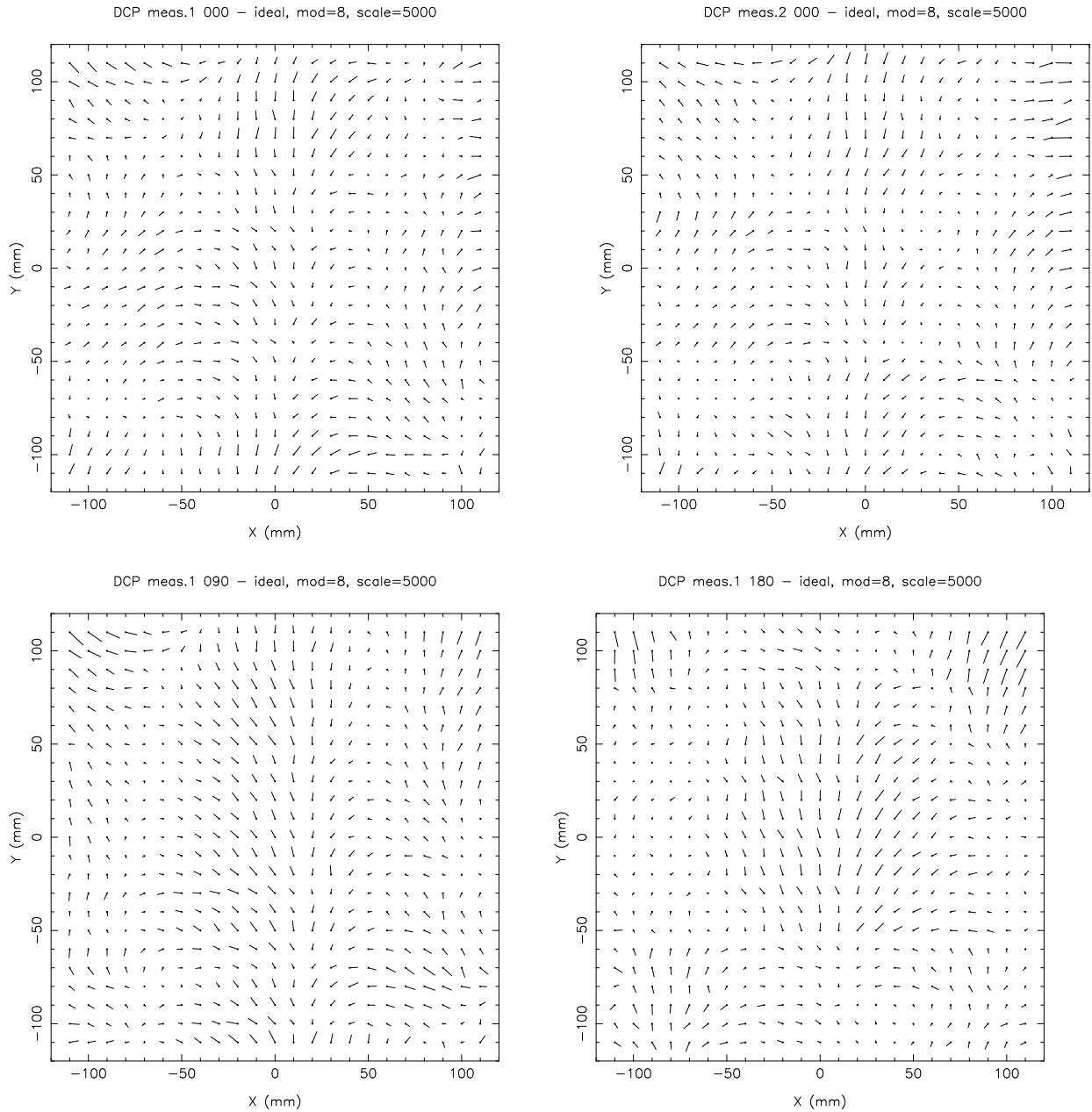

\caption{Differences between individual orientation measures
         of the dot calibration plate (DCP) and the ideal grid coordinates
         after a least-squares fit with an 8-parameter transformation
         model.
         The scale is 5000 thus the largest vectors are about 1.5 $\mu$m long.
         The top plots show results for the same orientation ($0^{\circ}$)
         from the first and second set. The plots at the bottom show
         more results from the first set for the $90^{\circ}$ and $180^{\circ}$
         orientation measures.}
\vspace*{5mm}
\plottwo{f5a.eps}{f5b.eps}
\vspace*{5mm}
\plottwo{f5c.eps}{f5d.eps}
\end{figure}

\begin{figure}
\caption{Mean StarScan $x,y$-table corrections.
         This pattern was derived by taking the smoothed average over all
         8 vector plots of which some are shown in Figure 4.
         The scale is 5000 thus the largest vectors are about 1.0 $\mu$m long.}
\vspace*{5mm}
\plotone{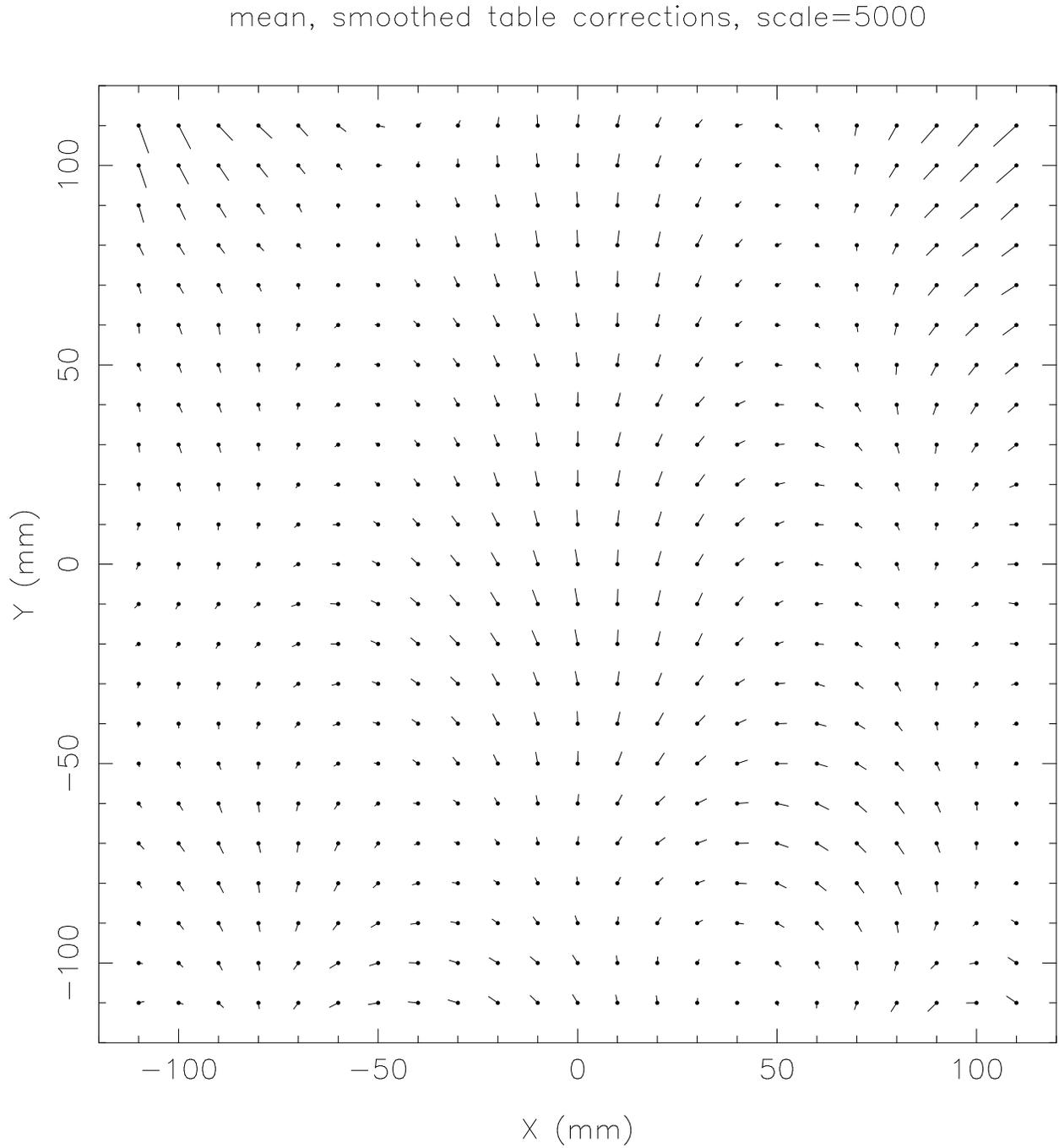}
\end{figure}

\begin{figure}
\caption{Data of Lick Astrograph plate 724 measured in 2 orientations
         in September 2004 (measure \# 6).
         The ``direct" minus ``reverse" residuals in $x$ are shown as
         a function of $x$, $y$, and magnitude (top to bottom).
         Each dot is the average over 25 stars, thus highlighting
         the systematic errors.}
\vspace*{5mm}
\epsscale{.7}
\plotone{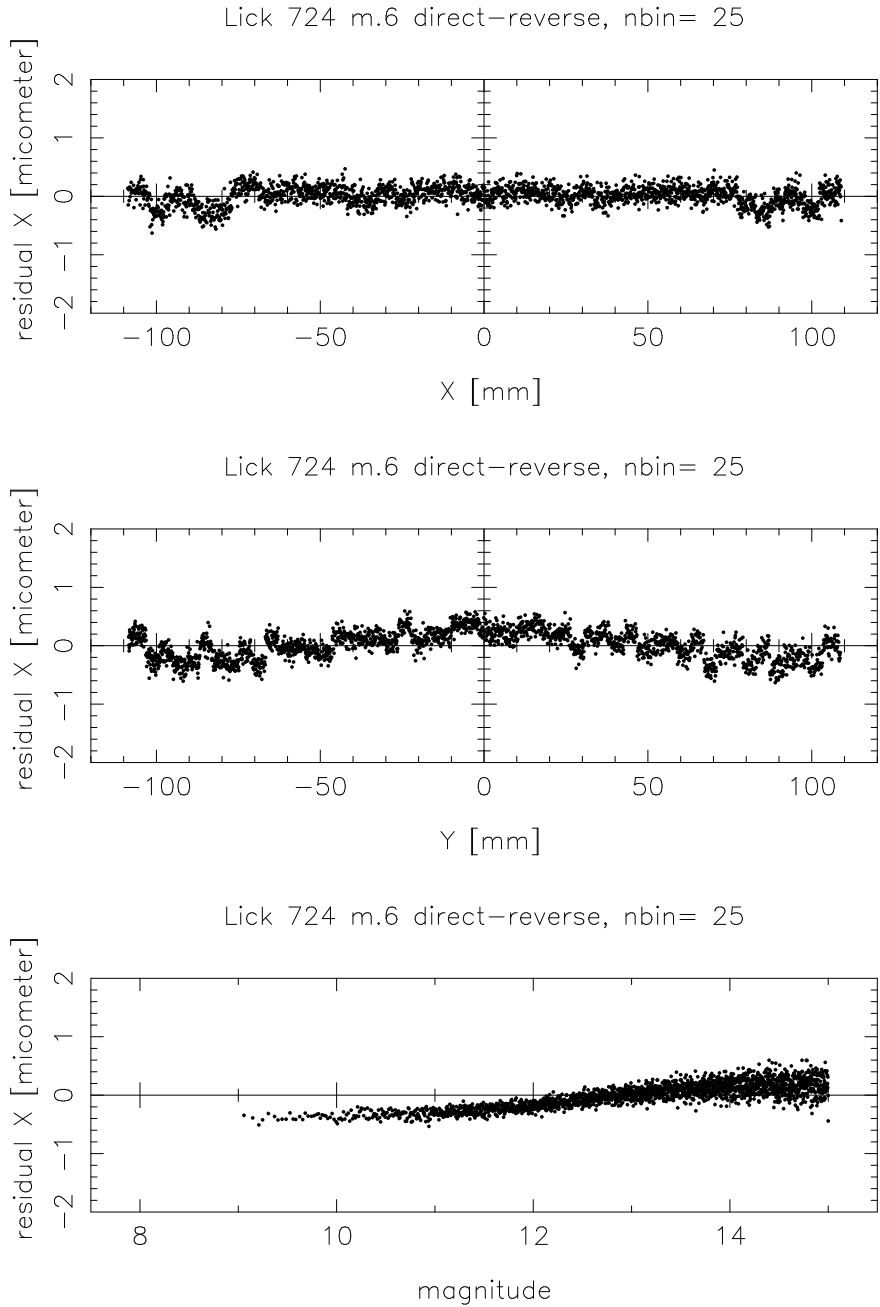}
\end{figure}

\begin{figure}
\caption{Data of Lick Astrograph plate 724 measured in 2 orientations.
         The ``direct" minus ``reverse" residuals in $x$ are shown as
         a function of $y$ for 3 different measure dates (from top
         to bottom):
         May 2004 (m.1), September 2005 (m.9), and April 2006 (m.a)
         Each dot is the average over 25 stars, thus highlighting
         the systematic errors.} 
\vspace*{5mm}
\plotone{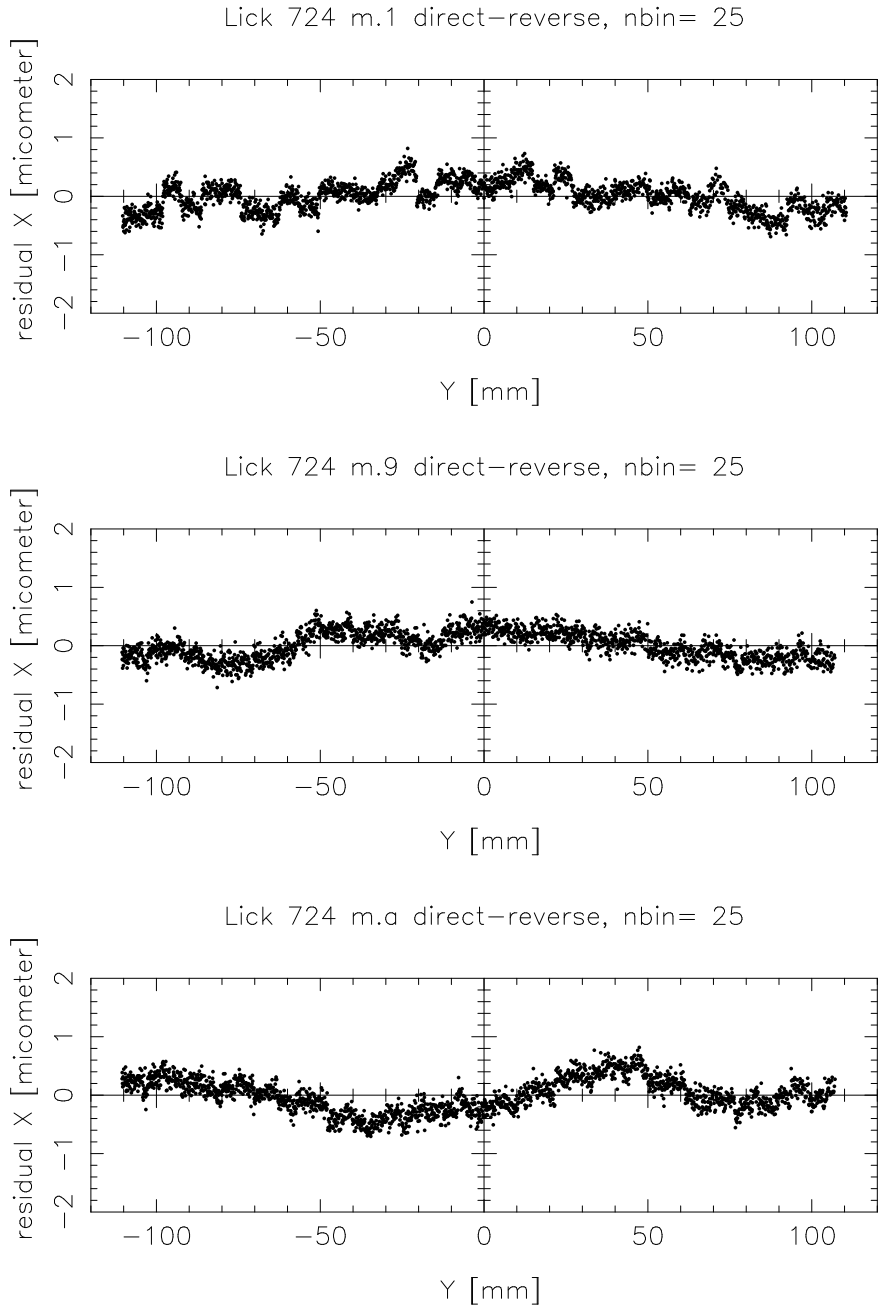}
\end{figure}

\end{document}